\documentclass[10pt,reqno,a4paper]{amsart}
\usepackage[british]{babel}
\usepackage{amssymb,amstext,amsthm,eucal}
\usepackage[utf8]{inputenc}
\usepackage{bbold}
\usepackage{array,color}
\usepackage[margin=2.8cm]{geometry}
\usepackage{enumitem}
\usepackage[small]{eulervm}
\usepackage{tgpagella}
\usepackage[unicode]{hyperref}
\hypersetup{%
  pdftitle   = {Eleven-dimensional supergravity from filtered
    subdeformations of the Poincaré superalgebra},
  pdfkeywords = {Lie superalgebras, Poincaré, supergravity,
    supersymmetry},
  pdfauthor  = {José Figueroa-O'Farrill, Andrea Santi},
  pdfcreator = {\LaTeX\ with package \flqq hyperref\frqq}
}
\theoremstyle{plain}

\newtheorem*{theorem*}{Theorem}
\theoremstyle{definition}

\newcommand{\ten}{\natural}

\newcommand{\Mat}{\operatorname{Mat}}
\newcommand{\AdS}{\operatorname{AdS}}

\newcommand{\GL}{\operatorname{GL}}
\newcommand{\SO}{\operatorname{SO}}
\newcommand{\CSO}{\operatorname{CSO}}

\newcommand{\Cl}{C\ell}

\newcommand{\stab}{\mathfrak{stab}}

\newcommand{\fso}{\mathfrak{so}}

\newcommand{\fm}{\mathfrak{m}}
\newcommand{\fp}{\mathfrak{p}}
\newcommand{\fs}{\mathfrak{s}}

\newcommand{\fh}{\mathfrak{h}}

\newcommand{\1}{\mathbb{1}}

\newcommand{\RR}{\mathbb{R}}
\newcommand{\ZZ}{\mathbb{Z}}

\newcommand{\be}{\boldsymbol{e}}
\newcommand{\sbar}{{\overline s}}

\newcommand{\gr}{\operatorname{gr}}

\allowdisplaybreaks
\begin{document}

\title{Eleven-dimensional supergravity from filtered subdeformations of the Poincaré superalgebra}
\author{José Figueroa-O'Farrill}
\author{Andrea Santi}
\address{Maxwell Institute and School of Mathematics, The University
  of Edinburgh, James Clerk Maxwell Building, Peter Guthrie Tait Road,
  Edinburgh EH9 3FD, United Kingdom}
\thanks{EMPG-15-23}
\begin{abstract}
We summarise recent results concerning the classification of filtered
deformations of graded subalgebras of the Poincaré superalgebra in
eleven dimensions, highlighting what could be considered a novel
Lie-algebraic derivation of eleven-dimensional supergravity.
\end{abstract}
\maketitle

\section{Introduction}
\label{sec:introduction}

In a recent paper \cite{JMFAS-Spencer} we have taken a first step in a
Lie-algebraic approach to the classification problem of supersymmetric
supergravity backgrounds.  In passing we provided what could be
considered a Lie-algebraic derivation of eleven-di\-men\-sional
supergravity.  We think that this result is interesting in its own
right, but it is hidden in the middle of a somewhat technical paper
and hence we fear that the message might be lost.  Therefore the
purpose of this short note is to summarise the results of
\cite{JMFAS-Spencer} and discuss their significance in as few words as
possible.  Readers who are pressed for time might consider reading
this short note before tackling the longer paper.

This note is organised as follows.  In
Section~\ref{sec:grad-subalg-poinc} we define the basic objects:
$\ZZ$-graded subalgebras of the eleven-di\-men\-sional Poincaré
superalgebra.  In Section~\ref{sec:filt-deform} we explain the problem
we set out to solve in \cite{JMFAS-Spencer}: namely, the
classification of filtered deformations of such subalgebras, and
summarise our results.  That is the content of the Theorem below.  In
Section~\ref{sec:spencer-cohomology} we mention one of the auxiliary
results obtained on the way to proving the Theorem: the calculation of
a Spencer cohomology group, which shows how the $4$-form emerges in
this approach.  Finally, in Section~\ref{sec:concluding-remarks} we
make some remarks about the results and try to discuss their
significance.

\section{Graded subalgebras of the Poincaré superalgebra}
\label{sec:grad-subalg-poinc}

We start with an eleven-di\-men\-sional lorentzian vector space
$(V,\eta)$, of ``mostly minus'' signature.  You may think of it as
$\RR^{1,10}$ with $\eta =
\operatorname{diag}(+1,\underbrace{-1,\cdots,-1}_{10})$.  We let
$\fso(V) \cong \fso(1,10)$ denote the Lorentz Lie algebra.  This
is the Lie algebra of $\eta$-antisymmetric linear transformations of
$V$; that is, $A \in \fso(V)$ if for all $v,w \in V$,
\begin{equation}
  \eta(Av,w) = - \eta(v,Aw)~.
\end{equation}
If we think of $(V,\eta)$ as a flat lorentzian manifold, the Lie
algebra of isometries is the Poincaré Lie algebra 
\begin{equation}
  \fp = \fso(V) \ltimes V~,
\end{equation}
where $V$ acts by infinitesimal translations.  The Lie brackets are
\begin{equation}
  \label{eq:poincare}
  [A,B] = AB - BA \qquad  [A,v] = Av \qquad\text{and}\qquad [v,w] = 0~,
\end{equation}
for all $A,B \in \fso(V)$ and $v,w\in V$.  If we grade $\fp$ by
declaring that $\fso(V)$ has degree $0$ and $V$ has degree\footnote{the
reason for this seemingly bizarre degree will become clear in a
moment} $-2$, we see that $\fp$ is a $\ZZ$-graded Lie algebra.

Let us now introduce supersymmetry.  Associated to $(V,\eta)$ we have
the Clifford algebra $\Cl(V,\eta) \cong \Cl(1,10)$.  Our conventions
for the Clifford algebra are such that
\begin{equation}
  v^2 = -\eta(v,v) \1 \qquad\forall v \in V~,
\end{equation}
and with our choice of $\eta$, we have an isomorphism of real associative algebras
\begin{equation}
  \Cl(V,\eta) \cong \Mat_{32}(\RR) \oplus \Mat_{32}(\RR)~,
\end{equation}
from where one reads that $\Cl(V,\eta)$ has precisely two inequivalent
irreducible Clifford modules $S_\pm$, which are real and of dimension
$32$.  They are distinguished by the action of the centre.  If we let
$\Gamma_{11} \in \Cl(V,\eta)$ denote the volume form, then
$\Gamma_{11}^2 = + \1$ and is central, so that it acts like a scalar
multiple of the identity on an irreducible Clifford module.  That
scalar multiple is $\pm 1$ on $S_\pm$.  We let $S = S_-$ from now on.

On $S$ we have a symplectic inner product $\left<-,-\right>$ which obeys
\begin{equation}
  \label{eq:symplectic}
  \left<v \cdot s_1, s_2\right> = - \left<s_1, v \cdot s_2\right>~,
\end{equation}
for all $v \in V$ and $s_1,s_2 \in S$ and where $\cdot$ stands for the
Clifford action.  It follows from equation~\eqref{eq:symplectic} that
$\fso(V)$ preserves $\left<-,-\right>$.  In fact, $S$ is a real
symplectic irreducible representation of $\fso(V)$: the spinor
representation.

The Poincaré algebra $\fp$ admits a supersymmetric extension: a
Lie superalgebra $\fs = \fs_{\bar 0} \oplus \fs_{\bar 1}$ with
$\fs_{\bar 0} = \fp$ and $\fs_{\bar 1} = S$.  The Lie brackets extend
the Lie brackets \eqref{eq:poincare} of $\fp$ by
\begin{equation}
  \label{eq:superpoincare}
  [A,s] = As \qquad [v,s] = 0 \qquad\text{and}\qquad [s_1,s_2] = \kappa(s_1,s_2)~,
\end{equation}
where $\kappa$, being symmetric, is determined by its value on the
diagonal, where it coincides with the Dirac current of $s \in S$,
defined by
\begin{equation}
  \label{eq:diraccurrent}
  \eta(\kappa(s,s),v) = \left<v\cdot s, s\right>~,
\end{equation}
for all $s \in S$ and $v \in V$.  Relative to an $\eta$-orthonormal
basis $\be_\mu = (\be_0,\be_1,\cdots,\be_9,\be_\ten)$ for $V$, and
writing
\begin{equation}
  \left<s, \be_\mu \cdot s \right> = \sbar \Gamma_\mu s~,
\end{equation}
we see that
\begin{equation}
  \kappa(s,s) = - \sbar \Gamma^\mu s \be_\mu~,
\end{equation}
where the minus sign comes from equation \eqref{eq:symplectic}.  The
Poincaré superalgebra $\fs$ is actually a $\ZZ$-graded super-extension
of the Poincaré algebra $\fp$: we simply let $S$ have degree $-1$ and
write
\begin{equation}
  \fs = \fs_0 \oplus \fs_{-1} \oplus \fs_{-2} = \fso(V) \oplus S \oplus V~.
\end{equation}
We remark that the $\ZZ$-grading and the parity are compatible in that
the parity is the degree modulo $2$.  This will be the case for all
the $\ZZ$-graded Lie superalgebras we shall consider.

Now let $\fh_0 < \fso(V)$ be any subalgebra of the Lorentz Lie
algebra, $S'\subset S$ be a subspace of $S$ which is stabilised (as a
subspace) by $\fh_0$, and let $V' \subset V$ be the subspace of $V$
spanned by the Dirac currents of all the spinors $s \in S'$.
Equivariance guarantees that $\fh_0$ also stabilises $V'$ as a
subspace.  Then $\fh = \fh_0 \oplus S' \oplus V'$ is a $\ZZ$-graded
subalgebra of the Poincaré superalgebra $\fs$ with respect to the
restriction to $\fh$ of the Lie brackets in
equations~\eqref{eq:poincare} and \eqref{eq:superpoincare}.  In this
note we will restrict to the special subalgebras with $S'=S$, so that
$V'=V$ (in fact, if $\dim S'> 16$ then $V'=V$ by the results of
\cite{HomogThm}).  Hence from now on we will consider $\ZZ$-graded 
subalgebras $\fh$ of $\fs$ which differ only in degree zero; that is,
\begin{equation}
  \label{eq:h}
  \fh = \fh_0 \oplus \fh_{-1} \oplus \fh_{-2} = \fh_0 \oplus S \oplus
  V\subset \fso(V) \oplus S \oplus V = \fs~.
\end{equation}

\section{Filtered deformations}
\label{sec:filt-deform}

Every $\ZZ$-graded Lie superalgebra admits a canonical filtration.  In the
case of the Lie superalgebra $\fh$ in equation~\eqref{eq:h}, this is a
filtration
\begin{equation}
  F^\bullet\fh : \qquad \fh = F^{-2}\fh \supset F^{-1}\fh \supset F^0\fh \supset 0~,
\end{equation}
where $F^{-1}\fh = \fh_{-1} \oplus \fh_0$, $F^0\fh = \fh_0$.  We can extend
it to $F^i\fh = \fh$ for all $i<-2$ and $F^i\fh = 0$ for all $i>0$.  The Lie
brackets are such that $[F^i\fh, F^j\fh] \subset F^{i+j}\fh$.

Now, any filtered Lie superalgebra $F^\bullet$ gives rise to an
associated graded Lie superalgebra $\gr_\bullet F$, defined by
\begin{equation}
  \gr_i F = F^i\bigr/F^{i+1}~,
\end{equation}
whose Lie brackets are the components of the Lie brackets of
$F^\bullet$ which have zero filtration degree.  Indeed, it follows
from the fact that $F$ is filtered, that
$[\gr_i F, \gr_j F] = \gr_{i+j}F$.  In the case of the filtered Lie
superalgebra $F^\bullet\fh$ associated to $\fh$ in
equation~\eqref{eq:h}, the associated graded $\gr_\bullet F\fh$ is
isomorphic to $\fh$ itself.

Of course there may be other filtered Lie superalgebras whose associated
graded algebra is again isomorphic to $\fh$.   They are called
\textbf{filtered deformations} of $\fh$.  We shall be concerned with
\textbf{nontrivial} filtered deformations: namely, those that are not
themselves isomorphic to $\fh$.

What does this mean in practice?  A filtered deformation of $\fh$ has
the same underlying vector space as $\fh$ but the Lie brackets are
obtained by adding to the Lie brackets of $\fh$ terms with
positive filtration degree.  Since the $\ZZ$-grading is compatible
with the parity, they always have even filtration degree.

For the case at hand, we can be much more explicit.  A filtered
deformation of $\fh$ in equation \eqref{eq:h} is given by
\begin{equation}
  \label{eq:fdef}
  \begin{aligned}[m]
    [A,B] &= AB - BA\\
    [A,s] &= As\\
    [A,v] &= Av + t \lambda(A,v)
  \end{aligned}
  \qquad\qquad
  \begin{aligned}[m]
    [s,s'] &= \kappa(s,s') + t \gamma(s,s')\\
    [v,s] &= t \beta(v,s)\\
    [v,w] &= t \tau(v,w) + t^2 \rho(v,w)~,
  \end{aligned}
\end{equation}
for all $A,B \in \fh_0$, $v,w \in V$ and $s,s' \in S$, where we
have introduced an inessential parameter $t$ to keep track of the
filtration degree and where the new brackets are
\begin{equation}
  \begin{aligned}[t]
    \lambda &: \fh_0 \otimes V \to \fh_0\\
    \tau &: \Lambda^2 V \to V\\
    \beta &: V \otimes S \to S\\
    \gamma &: \odot^2 S \to \fh_0
  \end{aligned}
  \qquad\qquad
  \begin{aligned}[t]
    \rho &: \Lambda^2 V \to \fh_0~.
  \end{aligned}
\end{equation}
Those in the first column have filtration degree $2$ and the one
in the second column has filtration degree $4$.

Of course, $\lambda,\tau,\rho,\beta,\gamma$ are not arbitrary, but
must satisfy the Jacobi identity (for all $t$).  In a recent paper
\cite{JMFAS-Spencer} we solved this problem and our solution is
contained in the Theorem below.  But first some notation.  By
$\CSO(V)$ we mean the Lie group of homothetic linear transformations
of $(V,\eta)$; that is, $\CSO(V) = \RR^\times \times \SO(V)$, acting
on $V$ by $v \mapsto \alpha L v$, where $\alpha \in \RR^\times$ and
$L \in \SO(V)$.  Recall, as well, that $\theta \in \Lambda^p V$ is
said to be \textbf{decomposable} if $\theta = v_1 \wedge \cdots \wedge
v_p$ for some $v_1,\cdots,v_p \in V$.

The main result of \cite{JMFAS-Spencer} can now be summarised as follows.

\begin{theorem*}
  The only subalgebras $\fh \subset \fs$ (in the class in
  equation~\eqref{eq:h}) which admit nontrivial filtered deformations
  are those for which $\fh_0 = \fso(V) \cap \stab(\varphi)$, where
  $\varphi \in \Lambda^4 V$ is nonzero and decomposable, and
  $\stab(\varphi)$ is the Lie algebra of its stabiliser in $\GL(V)$.
  In the nontrivial filtered deformations, the maps $\lambda$ and
  $\tau$ are zero and $\beta,\gamma,\rho$ are given explicitly in
  terms of $\varphi$ by
  \begin{equation}
    \label{eq:classi}
    \begin{split}
      \beta(v,s) &= v \cdot \varphi \cdot s - 3 \varphi \cdot v \cdot s\\
      \gamma(s,s) v &= - 2 \kappa(s, \beta(v,s))\\
      \rho(v,w) s &= [\beta_v, \beta_w]s~,
    \end{split}
  \end{equation}
  where $\beta_v s= \beta(v,s)$.  Moreover, if $\varphi, \varphi' \in
  \Lambda^4 V$ are decomposable and in the same orbit of $\CSO(V)$,
  then the resulting filtered deformations are isomorphic.
\end{theorem*}

Notice that from equation \eqref{eq:classi} it follows that we can
reabsorb the parameter $t$ in equation \eqref{eq:fdef} into $\varphi$,
since both $\beta$ and $\gamma$ are linear in $\varphi$, whereas
$\rho$ is quadratic in $\varphi$.

There are precisely three $\CSO(V)$-orbits of decomposable nonzero
$4$-forms in $\Lambda^4 V$.  Every nonzero decomposable
$\varphi = v_1 \wedge v_2 \wedge v_3 \wedge v_4$ defines a $4$-plane
$\Pi \subset V$ --- namely, the real span of the $(v_i)$ --- and
conversely, every $4$ plane $\Pi \subset V$ gives rise to an
$\RR^\times$ orbit of decomposable $\varphi$ by taking the wedge
product of any basis.  Two different bases will give $\varphi$'s which
are related by the determinant of the change of basis, so in the same
$\RR^\times$ orbit.

The inner product $\eta$ restricts to $\Pi$ in one of three ways:
\begin{enumerate}[label=(\alph*)]
\item $(\Pi,\eta\bigr|_\Pi)$ is a lorentzian $4$-plane,
\item $(\Pi,\eta\bigr|_\Pi)$ is a euclidean $4$-plane, or
\item $(\Pi,\eta\bigr|_\Pi)$ is a degenerate plane.
\end{enumerate}
We can choose representative $\varphi$'s for each case, respectively:
\begin{enumerate}[label=(\alph*)]
\item $\varphi = \be_0 \wedge \be_1 \wedge \be_2 \wedge \be_3$,
\item $\varphi = \be_7 \wedge \be_8 \wedge \be_9 \wedge \be_\ten$, or
\item $\varphi = \be_+ \wedge \be_1 \wedge \be_2 \wedge \be_3$,
\end{enumerate}
where $\be_+ = \be_0 + \be_\ten$.
The resulting filtered Lie superalgebras are precisely the Killing
superalgebras of the maximally supersymmetric backgrounds of
eleven-di\-men\-sional supergravity, respectively:
\begin{enumerate}[label=(\alph*)]
\item the family of Freund--Rubin backgrounds $\AdS_4 \times S^7$ \cite{FreundRubin},
\item the family of Freund--Rubin backgrounds $S^4 \times \AdS_7$ \cite{AdS7S4}, or
\item the Kowalski-Glikman pp-wave \cite{KG}.
\end{enumerate}

\section{Spencer cohomology}
\label{sec:spencer-cohomology}

The proof of the Theorem, broken down into a number of preliminary
results, occupies most of \cite{JMFAS-Spencer} and we invite those
interested to read the details of the proof in that paper.  That being
said, it might be worth spending a few words on one aspect of the
calculation, since it is this aspect which justifies the title of this
note.

Like almost every other algebraic deformation problem, the
classification of filtered deformations is governed by an underlying
cohomology theory.  In this case, it is Spencer cohomology, which is
intimately linked to the Chevalley--Eilenberg
\cite{ChevalleyEilenberg} cohomology of the supertranslation ideal
$\fm$ of $\fh$ with values in $\fh$ itself viewed as an $\fm$-module
by restricting the adjoint representation.   Let us be more precise.

Let $\fm = \fh_{-1} \oplus \fh_{-2} = S \oplus V$ denote the
supertranslation ideal of $\fh$.  (Notice that $\fh \subset \fs$ could
be taken to be $\fs$ itself.)  Restricting the
adjoint representation of $\fh$ to $\fm$, $\fh$ becomes an
$\fm$-module and we may consider the Chevalley--Eilenberg complex
$C^\bullet(\fm;\fh)$.  Because the Lie superalgebra $\fh$ is
$\ZZ$-graded, the differential has zero degree and hence
$C^\bullet(\fm;\fh)$ admits a decomposition into subcomplexes
$C^{\bullet,\bullet}(\fm;\fh)$ labelled by the degree.  The subcomplex
$C^{d,\bullet}(\fm;\fh)$ is the Spencer complex of degree $d$ and its
cohomology $H^{d,\bullet}(\fm;\fh)$ is the Spencer cohomology of
degree $d$.  Furthermore, the Spencer differential is
$\fh_0$-equivariant and hence the Spencer cohomology is naturally an
$\fh_0$-module.

Infinitesimal filtered deformations are classified by the Spencer
cohomology group $H^{2,2}(\fm;\fh)$.  In \cite{JMFAS-Spencer} we
compute this group by bootstrapping the simpler calculation of
$H^{2,2}(\fm;\fs)$, which yields an $\fso(V)$-module isomorphism
\begin{equation}
  H^{2,2}(\fm;\fs) \cong \Lambda^4 V~,
\end{equation}
with the cocycle corresponding to $\varphi \in \Lambda^4 V$ given by
$\beta \oplus \gamma$ in equation~\eqref{eq:classi}.

In particular, we see that the $4$-form \emph{emerges} from the
calculation.  If we put $\varphi = \tfrac1{24} F$, for $F$ the
$4$-form flux in eleven-di\-men\-sional supergravity, we recognise at once
the cochain $\beta$ as the zeroth order term in the gravitino
connection.  In other words, we have in effect \emph{derived} the
gravitino connection
\begin{equation}
  D_v s = \nabla_v s - \beta_v s = \nabla_v s - v \cdot \varphi \cdot
  s + 3 \varphi \cdot v \cdot s
\end{equation}
from the Spencer cohomology of the Poincaré superalgebra.  The
gravitino connection encodes all the information necessary to define
the notion of a supersymmetric background.  Indeed, a background is
one for which the gamma-trace of the curvature of $D$ vanishes
\cite{GauPak} and it is supersymmetric precisely when $D$ has nonzero
kernel. Therefore one might be so bold as to say that, at least
insofar as the bosonic backgrounds of the theory are concerned, we
have re-derived eleven-dimensional supergravity.  Of course, to
actually re-derive the supergravity action, requires climbing
the same mountain that Cremmer, Julia and Scherk did in \cite{CJS},
albeit departing from slightly higher ground.

\section{Concluding remarks}
\label{sec:concluding-remarks}

It is possible to show that the Killing superalgebra (KSA) of
\emph{any} supersymmetric background of eleven-di\-men\-sional
supergravity is a filtered deformation of a $\ZZ$-graded subalgebra of
the Poincaré superalgebra $\fs$.  (The proof will appear in a
forthcoming paper.)  It is thus not surprising that we should find the
KSAs of the maximally supersymmetric backgrounds among the filtered
deformations of a subalgebra of the form $\fh$ in
equation~\eqref{eq:h}.  However it is important to emphasise that we
did \emph{not} set out to classify KSAs.  That is a different problem,
since the maps $\lambda, \tau, \rho, \beta, \gamma$ are then already
partially determined \emph{a priori}.  So the surprising fact, albeit
growing less surprising in the telling, is that we find \emph{nothing
  else} but the KSAs.

Our derivation in \cite{JMFAS-Spencer} of the maximally supersymmetric
backgrounds is to be contrasted with the standard derivation.  This
starts from the same initial data: namely, the Poincaré superalgebra
$\fs$.  Then one first finds evidence, again purely by
representation-theoretic arguments, of the existence of a supergravity
theory in eleven dimensions \cite{Nahm}, one then constructs the
supergravity theory \cite{CJS} (which includes discovering $D$) and
then solves the $D$-flatness equations \cite{FOPMax}.  Compared to the
classical route to the result, one could perhaps argue that our recent
approach enjoys a certain conceptual economy.

Finally, although we restricted ourselves in \cite{JMFAS-Spencer} to
eleven-di\-men\-sional supergravity, we have done similar calculations in
other dimensions, which will be the subject of forthcoming papers.  In
particular, in collaboration with Paul~de~Medeiros, we have performed
this analysis for $N=1$ $d=4$ supergravity and can reproduce the
Killing spinor equations in \cite{Festuccia:2011ws}.  Our results have
appeared in \cite{deMedeiros:2016srz}.

\section*{Acknowledgments}

The research of JMF is supported in part by the grant ST/L000458/1
``Particle Theory at the Tait Institute'' from the UK Science and
Technology Facilities Council, and that of AS is fully supported by a
Marie-Curie research fellowship of the ''Istituto Nazionale di Alta
Matematica'' (Italy).  We are grateful to our respective funding
agencies for their support.  This note was written while JMF
was a guest of the Grupo de Investigación ``Geometría diferencial y
convexa'' of the Universidad de Murcia, and he wishes to thank Ángel
Ferrández Izquierdo for the invitation, the hospitality and the fine
weather.

\bibliographystyle{utphys}
\bibliography{JPhysA}

\end{document}